\documentclass[aps,prl,twocolumn,groupedaddress]{revtex4-1}%
\usepackage{amsfonts}
\usepackage{amsmath}
\usepackage{amssymb}
\usepackage{graphicx}%
\providecommand{\U}[1]{\protect\rule{.1in}{.1in}}
\providecommand{\U}[1]{\protect\rule{.1in}{.1in}}
\providecommand{\U}[1]{\protect\rule{.1in}{.1in}}

\begin{document}
\title{Critical-doping universality for cuprate superconductors: Oxygen
nuclear-magnetic-resonance investigation of (Ca$_{x}$La$_{1-x}$)(Ba$_{1.75-x}%
$La$_{0.25+x}$)Cu$_{3}$O$_{y}$}
\author{Eran Amit}
\author{Amit Keren}
\affiliation{Physics Department, Technion, Israel Institute of Technology, Haifa 32000, Israel}
\pacs{PACS number}

\begin{abstract}
The critical doping levels in cuprates, where the ground state changes its
nature (from an antiferromagnet to a spin glass to superconductor to metal),
are not universal. We investigate the origin of these critical doping
variations by measuring the in-plane oxygen $p_{\sigma}$ hole density in the
CuO$_{2}$ layers as a function of the oxygen density $y$ in (Ca$_{x}$%
La$_{1-x}$)(Ba$_{1.75-x}$La$_{0.25+x}$)Cu$_{3}$O$_{y}$. This is done using the
oxygen 17 nuclear quadrupole resonance parameter $\nu_{Q}$. We compare
compounds with $x=0.1$ and $0.4$ which have significant critical $y$
variations and find that these variations can be explained by a change in the
efficiency of hole injection into the $p_{\sigma}$ orbital. This allows us to
generate a unified phase diagram for the CLBLCO system across the entire
doping range, with no adjustable parameters.

\end{abstract}
\maketitle

\begin{figure}
[ptb]
\begin{center}
\includegraphics[
trim=0.000000in 0.139317in 0.000000in 0.186062in,
natheight=4.582800in,
natwidth=3.080100in,
height=3.6463in,
width=2.6476in
]%
{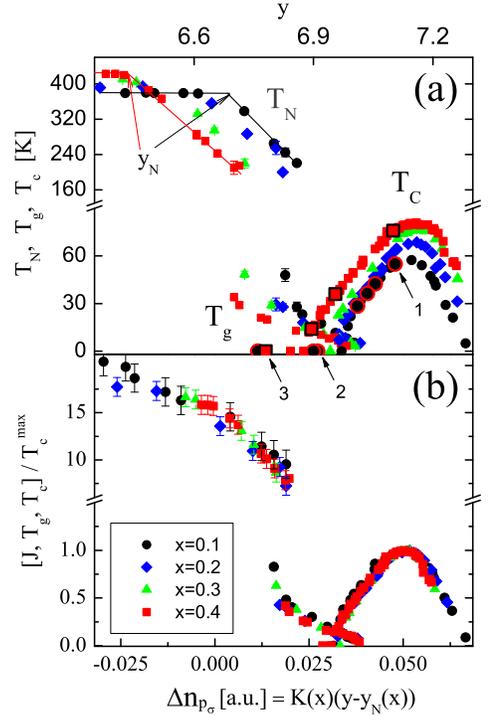}%
\caption{(color online) (Ca$_{x}$La$_{1-x}$)(Ba$_{1.75-x}$La$_{0.25+x}%
$)Cu$_{3}$O$_{y}$ phase diagram (a) showing the superconductivity, Ne\'{e}l,
and glass critical temperatures $T_{c}$, $T_{N}$ and $T_{g}$ respectivly as a
function of $x$ and $y$ \cite{ROfer}. The arrows mark the oxygen densities where $T_{N}$
starts to decrease. Fig. (b) is obtained from (a) as described in the text.}%
\label{clblco phase diagram}%
\end{center}
\end{figure}

There are several critical doping levels in the phase diagram of the cuprates
where the ground state changes \cite{Lee}. The first critical doping level is when the
long range antiferromagnetic (AFM) order is destroyed and replaced by a spin
glass (SG) state; next superconductivity (SC) emerges; then the spin glass is
destroyed; and finally, superconductivity is destroyed and replaced by a
metallic state. These critical levels exist in the phase diagram of all
cuprates which can be doped over a wide range such as La$_{2-x}$Sr$_{x}%
$CuO$_{4}$ (LSCO) and YBa$_{2}$Cu$_{3}$O$_{y}$ (YBCO), but they vary between
compounds. Several attempts have been made to construct a universal phase
diagram but thus far only partial diagrams, of only one or two phases, have
been achieved \cite{Uemura, Homes, Honma, Mourachkine, Zhou}. One particular
example is the phase diagram of the (Ca$_{x}$La$_{1-x}$)(Ba$_{1.75-x}%
$La$_{0.25+x}$)Cu$_{3}$O$_{y}$ (CLBLCO) system shown in
Fig.~\ref{clblco phase diagram}(a), which includes four different families
with $x=0.1,0.2,0.3$ and $0.4$ \cite{ROfer}. This phase diagram clearly demonstrates that the critical oxygen densities $y$ depend on $x$ and thus are not universal.

The reason for lack of universality is not clear and could be one of many. For
example, it is possible that the doping efficiency of the CuO$_{2}$ planes is
family-dependent. A second option is interlayer coupling, which in CLBLCO is
$x$ dependent \cite{ROfer}; it is conceivable that the interlayer coupling
determines the critical doping. Another possibility is that two different
kinds of holes are formed in the CuO$_{2}$ planes, and only the "mobile holes"
participate in the SC mechanism; perhaps the level of mobility varies between
families \cite{Kanigel, Sanna}. Finally, in the $t-J$ model the critical
doping where the AFM order is destroyed depends on $t/J$ \cite{Calandra}; it
could be that $t/J$ varies between CLBLCO families. In this work we
investigate the origin of the critical doping level variation between
different CLBLCO families by directly measuring the hole density in the
CuO$_{2}$ plane of CLBLCO using the oxygen Nuclear Quadrupole Resonance (NQR)
parameter $^{17}\nu_{Q}$. This parameter is extracted from Nuclear Magnetic
Resonance (NMR) experiments and is directly related to the density of holes in
the in-plane oxygen $p_{\sigma}$ orbital as we demonstrate below.

%

Our investigation concentrates on CLBLCO since its phase diagram is smooth and
systematic. It also does not have abrupt features such as the kink in the LSCO
SC dome at $x=0.125$ \cite{Takagi}, or a structural phase transition and chain ordering as
in YBCO. In fact, CLBLCO has a tetragonal structure (similar to underdoped
YBCO) for all values of $x$ and $y$. Consequently, the layer equivalent to the
"chain layer" of the YBCO can hold oxygen in both the $a$ and $b$ directions.
Therefore, for each family, the parameter $y$ varies between $6.4$ and $7.25$
and controls the doping level. The crystal quality and the total cation charge
are also $x$-independent \cite{Keren}. These properties reduces the number of
variable determining $T_{c}$. Thus, understanding superconductivity in this
compound could shed light on all other cuprates.

As $x$ increases from $x=0.1$ to $x=0.4$ the maximum $T_{c}$ ($T_{c}^{max}$)
increases from 57 to 81~K; the glass temperature $T_{g}$ decreases; and the
Ne\'{e}l temperature ($T_{N}$) of the parent compound increases from 380 to
425~K. Another important feature of CLBLCO is that for each family there is an
oxygen density which marks a transition between a constant $T_{N}$ and a
decreasing $T_{N}$ as a function of $y$, as demonstrated in
Fig.~\ref{clblco phase diagram}(a). We denote this density as $y_{N}$; for the
$x=0.1,0.2,0.3,0.4$ families $y_{N}=6.69,6.63,6.52,6.43,$ respectively.

Due to the systematic behavior of CLBLCO, the four different phase diagrams in
Fig.~\ref{clblco phase diagram}(a) can be reduced into one unified diagram in
three steps \cite{ROfer,Kanigeltg}: 1. Extracting the values of the in-plane AFM
coupling $J$ from $T_{N}$ for each family by dividing out the interplane
coupling contribution. 2. Dividing $J$, $T_{g}$, and $T_{c}$ of each family by
$T_{c}^{max}$ of that family. 3. Stretching the oxygen density axis for each
family around its $y_{N}$ by a factor $K(x)$, namely, introducing the quantity
$K(x)(y-y_{N})$, where $K=0.113,0.098,0.079,0.069$, for the $x=0.1$ to $0.4$
families respectively. The resulted scaled phase diagram is shown in Fig.
\ref{clblco phase diagram}(b). The scaling procedure is somewhat different
from that in previous works where the stretching was done around the oxygen
density of $T_{c}^{max}$ and a different set of $K$s was used
\cite{Keren,ROfer, ROfer2, Luba}. Multiplying all the $K$s by a numerical
factor would yield equally good data collapse. Here the $K$s are chosen so
that optimal doping is at $K(x)(y-y_{N})=0.05$, for reasons that will become
clear below. The ratio of $K$ between the $x=0.4$ and $x=0.1$ families is
$1:1.62$. Using the scaling terminology the question we address experimentally
in this work is: can $K(x)$ be explained by the in-plane oxygen $p_{\sigma}$
orbital doping efficiency, or is some more exotic explanation required?

\begin{figure}
[ptb]
\begin{center}
\includegraphics[
trim=0.099987in 0.166001in 0.221592in 0.221667in,
height=2.3705in,
width=3.3754in
]%
{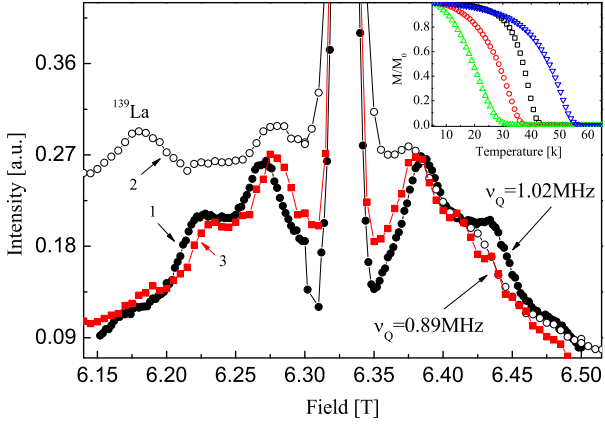}%
\caption{(color online) Raw NMR data of three samples marked in
Fig~\ref{clblco phase diagram}(a): 1 (closed circles) is x=0.1 close to
optimal doping (y=7.105), 2 (open circles) is x=0.1 underdoped (y=6.9), 3
(closed squares) is x=0.4 underdoped (y=6.79). From the high fields data it is
clear that the quadrupole frequency of samples 2 and 3 are almost identical
but different from sample 1. The dominant contribution in sample 2 at lower
fields is from the $^{139}$La nuclei. The inset shows the temperature
dependence of the normalized magnetic moment for the x=0.1 SC samples with
y=7.105 (up triangles), 7.055 (circles), 7.035 (squares) and 7.01 (down
triangles).}%
\label{compare}%
\end{center}
\end{figure}

For our experiments, sintered pellets of CLBLCO with different $x$ values were
prepared using standard techniques~\cite{Goldschmidt}, and then ground into
powder. Since only $^{17}$O has nuclear spin but its natural abundance is only
$0.038\%$, the samples were enriched with this isotope. In the enrichment
process the samples were heated to $520^{\circ}\mathrm{C}$ for five days and
then cooled to $320^{\circ}\mathrm{C}$ for five more days in enriched oxygen
gas with an isotope fraction of $40-50\%$. In order to obtain different oxygen
levels some of the enriched samples were later annealed in either natural
oxygen or nitrogen environments at different temperatures for 24 hours, and
then quenched in liquid nitrogen. When possible, $T_{c}$ was determined with a
magnetometer as shown in the inset of Fig. \ref{compare}. We used cryogenic
SQUID, at a field of 5Oe in FC conditions. In this inset the
normalized magnetization of four samples with $x=0.1$ and
$y=7.105,7.055,7.035,7.01$ is shown, demonstrating the variation in $T_{c}$.
The oxygen density was obtained from the known CLBLCO phase diagram
(Fig.~\ref{clblco phase diagram}). For the very underdoped samples, which are
not superconducting (see samples 1 and 2 in Fig. \ref{clblco phase diagram}),
the oxygen density was determined by iodometric titration \cite{Appelman}.

In our NMR experiment all the eleven samples emphasized in Fig.
\ref{clblco phase diagram}(a) by enlarged symbols were measured. We used
constant frequency $f=36.525$~MHz and temperature of $110$~K (the dependence
of $\nu_{Q}$ on temperature is within our experimental error \cite{Grevin}).
Measurements were performed over a range of external magnetic fields. For each
field 20480 spin echo sequences were collected. The intensity for each field
is the integral over the Fourier transform of the raw data.

There are three oxygen sites in CLBLCO: the planar O$_{23}$, apical O$_{4}$
and "chain" O$_{1}$. The NMR line of the O$_{1}$ site is negligible
\cite{Oldfield}. The apical oxygen does not affect the measured NMR lineshape
either. We measure at $110$~K, where its intensity is much smaller than the
planar oxygen and its line is wider \cite{Oldfield}. As a result, the spectrum
is dominated by only one oxygen nucleus instead of three. However, the
$^{139}$La nucleus appears in some of the measurements when the amount of
$^{17}$O in the sample is small. In our working frequency the $^{139}$La has a
central transition at $6.1~$T, while the $^{17}$O central transition is at
$6.33$~T. Therefore, in samples with a low enough concentration of $^{17}$O
the lanthanum signal changes the NMR line shape at low fields, but not at high
fields.%

We present in Fig. \ref{compare} NMR lines of three samples: (1) $x=0.1$ close
to optimal doping, (2) $x=0.1$ underdoped, and (3) $x=0.4$ very under doped.
Their place on the phase diagram is indicated in Fig.
\ref{clblco phase diagram}(a). There is a clear difference around $6.43$~T
between lines (1) and (3) but lines (2) and (3) are similar. As we explain
below, this is a consequence of different $p_{\sigma}$ densities in samples
(1) and (3), and similar densities in samples (2) and (3). In the lower field
regime of the underdoped sample (2) the lanthanum signal dominates the spectrum.

A quadrupole nucleus such as $^{17}$O can be viewed as a non spherical charge
distribution whose energy depends on its orientation with respect to the
internal electric fields. The nuclear Hamiltonian in an external magnetic
field $\mathbf{H}$ is a sum of the usual Zeeman interaction and additional
quadrupole term, and is given by%
\begin{equation}
\mathcal{H}=-\gamma\hbar\mathbf{H(1+\sigma)I}+\frac{eQV_{zz}}{4I\left(
2I-1\right)  }\left[  3I_{z}^{2}-I^{2}+\eta\left(  I_{x}^{2}-I_{y}^{2}\right)
\right]  \label{NQR Hamiltonian}%
\end{equation}
\noindent where $^{17}\gamma=5.77$~MHz/T is the gyromagnetic ratio,
$\mathbf{I}$ is the nuclear spin operator in the $I=5/2$ representation,
$\mathbf{\sigma}$ is the shift tensor and $eQ$ is the $^{17}$O quadrupole
moment. The second term is written in the Electric Field Gradient (EFG) system
representation where the EFG tensor is diagonal, $V_{zz}=\frac{\partial^{2}%
V}{\partial z^{2}}$ is the largest (axial) EFG eigenvalue, and $\eta
=(V_{xx}-V_{yy})/V_{zz}$ is the orthorhombic EFG asymmetry.

In the limit of small quadrupole frequency compared to $\gamma\hbar H$, the
energy difference between two nuclear spin states $\Delta E_{m\rightarrow
m-1}$ is given by~\cite{Baugher}:
\begin{align}
\Delta E_{m\rightarrow m-1}  &  ={\gamma hH}[1-\sigma_{i}]-h\nu_{Q}[\frac
{1}{2}(3\cos^{2}(\theta)-1)\label{energy difference}\\
&  -\frac{1}{2}\eta\sin^{2}(\theta)\cos(2\phi)](m-\frac{1}{2})\nonumber
\end{align}
where $m$ is the nuclear spin component parallel to the external (Zeeman)
magnetic field, $\sigma_{i}=\sigma_{x}+\sigma_{y}+\sigma_{z}$ is the diagonal
term of the chemical and Knight shifts (we neglected the off-diagonal terms),
and $\theta$ and $\phi$ are the angles between $\mathbf{z}$ and the external
magnetic field, and the quadrupole frequency is defined as%

\begin{equation}
\nu_{Q}=\frac{eQV_{zz}}{4I\left(  2I-1\right)  } \label{NuQ def}%
\end{equation}

\begin{figure}
[ptb]
\begin{center}
\includegraphics[
trim=0.139415in 0.185704in 0.232502in 0.232705in,
height=2.6109in,
width=3.5535in
]%
{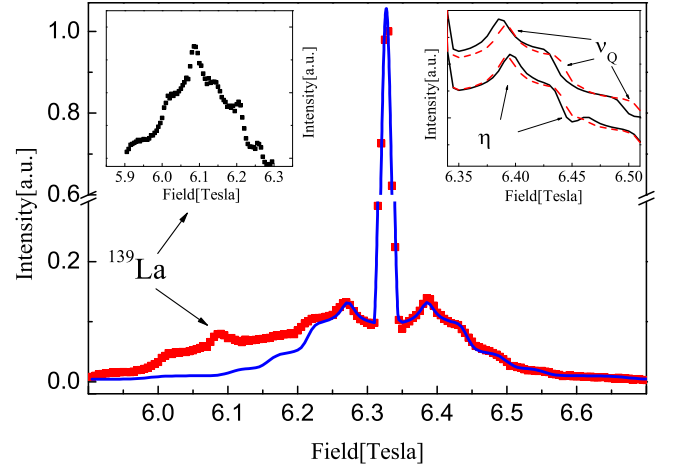}%
\caption{(color online) Raw data of an NMR measurement (red squares) and a fit
of the high field data to Eq.~\ref{fit integral} (blue line). At low fields
there is a deviation from the raw data due to the $^{139}$La line. The
difference between the raw data and the fit is presented in the left inset.
The right inset contains theoretical plots of samples with two different
$\nu_{Q}$ (top lines) and two different $\eta$ (bottom lines), as explained in
the text.}%
\label{oxygen nmr}%
\end{center}
\end{figure}

A resonance occurs when the frequency $f_{m}\left(  H,\nu_{Q},\sigma_{i}%
,\eta,\phi,\theta\right)  =\Delta E_{m\rightarrow m-1}/h$ equals an applied
frequency $f$. In powder, all possible orientations and line broadening must
be taken into account. Therefore, the spectrum is given by%
\begin{align}
I(H) &  =%
{\displaystyle\sum\limits_{m=-3/2}^{5/2}}
W\left(  m\right)  \underset{0}{\overset{\infty}{\int}}d\sigma_{i}^{\prime
}e^{-\frac{\left(  \sigma_{i}^{\prime}-\sigma_{i}\right)  ^{2}}{2\Delta
\sigma_{i}^{2}}}\underset{0}{\overset{\infty}{\int}}d\nu_{Q}^{\prime}%
e^{-\frac{\left(  \nu_{Q}^{\prime}-\nu_{Q}\right)  ^{2}}{2\Delta\nu_{Q}^{2}}%
}\label{fit integral}\\
&  \int d\Omega\delta\left(  f_{m}\left(  H,\nu_{Q}^{\prime},\sigma
_{i}^{\prime},\eta,\phi,\theta\right)  -f\right)  .\nonumber
\end{align}
The line broadening in our experiments has typical values of $\Delta\nu
_{Q}<0.2\nu_{Q}$ and $\Delta\sigma_{i}\cong0.002\sigma_{i}$. Finally $W(m)$
represents the weights of the different transitions and is taken as fit
parameter. The effect of the parameter on the line shape is demonstrated in
the right panel of Fig. \ref{oxygen nmr}. The high field side of the
theoretical lines for $\nu_{Q}=0.95$, $1.03$~MHz (with $\eta=0.33$) and of
$\eta=0.28$ ,$0.32$ (with $\nu_{Q}=1.02$ MHz) are plotted in the right inset.
Arrows mark the regions in which $\eta$ and $\nu_{Q}$ changes have the most
effect on the spectrum, and enable us to distinguish between these parameters.%

\begin{figure}
[ptb]
\begin{center}
\includegraphics[
trim=0.105811in 0.158888in 0.212485in 0.265377in,
height=2.4815in,
width=3.3508in
]%
{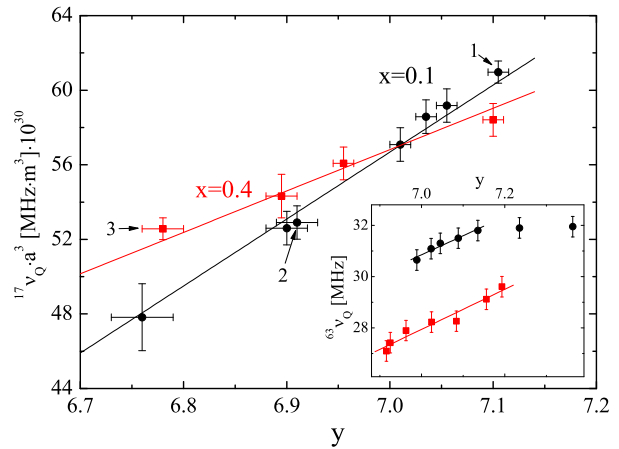}%
\caption{(color online) The translation of the number of oxygen atoms per unit
cell (abscissa) into the number of $p_{\sigma}$ holes (ordinate), as extracted
from the NMR data. The x=0.1 family is in black circles, and the x=0.4 family
is in red squares. The ratio between the slopes is equal to the stretching
ratio between the families in the scaling process shown in Fig.
\ref{clblco phase diagram}\ (see text). The place on the phase diagram of
samples 1 to 3 is shown in Fig.~\ref{clblco phase diagram}(a), and their raw
NMR data is depicted in Fig.~\ref{compare}. The inset shows that measurements
of the copper quadrupole frequencies in CLBLCO give the same slope for both
families \cite{Kanigel}.}%
\label{slopes}%
\end{center}
\end{figure}

An NMR line of the CLBLCO sample with $x=0.4$ and $y=7.1$ (solid symbols) is
presented in the main panel of Fig. \ref{oxygen nmr}. The solid line is the best fit of Eq. \ref{fit integral} to the data at the high field side. It gives $\nu
_{Q}=0.98~$MHz and $\eta=0.33$. These numbers are similar to previous
measurements of YBCO \cite{Oldfield}. The difference between the fit and the
data on the low field side, which is caused by the lanthanum nuclei, is
plotted in the left inset. In order to obtain $\nu_{Q}(x,y)$ from all samples
we fit Eq. \ref{fit integral} to the NMR spectrum of all the measured samples.
The results are shown in Fig.~\ref{slopes}.

The EFG on a planar oxygen site is induced by the electrons and nuclei
surrounding the oxygen \cite{Slichter}. The principal axis of this EFG ($z$
direction) is parallel to the copper-oxygen-copper axis \cite{Oldfield}. The
two main contributions to $V_{zz}$ are : I) holes in the oxygen $p_{\sigma}$
orbital; II) holes and nuclei of the atoms surrounding the oxygen. The first
contribution is directly proportional to the number of holes created by the
doping process. In contrast, (II) has a negligible dependence on doping
\cite{Hasse}. Moreover, the holes in the oxygen $p_{\sigma}$ are much closer
to the nucleus and therefore their contribution to the EFG is more significant.

The classical formula of $V_{zz}$ is given by $V_{zz}=\int
\rho(\overrightarrow{r})\frac{r^{2}-3z^{2}}{r^{5}}d^{3}r$ where $\rho$ and $r$
are the charge density and distance from the nucleus, respectively. The value
of $V_{zz}^{\prime}$ induced by a different charge distribution given by
$\rho^{\prime}(\overrightarrow{r})=\xi^3\rho(\xi\overrightarrow{r})$, where $\xi$
is a constant, is
\begin{equation}
V_{zz}^{\prime}=\xi^{3}V_{zz}. \label{quad point}%
\end{equation}

The oxygen $p_{\sigma}$ electronic wave functions of the different CLBLCO
samples differ mostly in their typical length-scale. The characteristic length
is proportional to the unit cell parameter $a$. Neutron diffraction
experiments show that $a$ changes only by about one percent between the
different families, and by 0.1 percent within a family \cite{ROfer2}.
Therefore, using Eq. \ref{quad point} and Eq. \ref{NuQ def} it is expected
that $\Delta(V_{zz}a^{3})\propto\Delta(\nu_{Q}a^{3})\propto\Delta
n_{p_{\sigma}}$, where $n_{p_{\sigma}}$ is the hole density in the oxygen
$p_{\sigma}$ orbital and $\Delta$ stands for the changes induced by the doping process.

In Fig. \ref{slopes} we present $\nu_{Q}a^{3}$ versus oxygen levels $y$ for
the two families with $x=0.1$ and $0.4$ (Samples 1, 2 and 3 are the same as in
Fig.~\ref{clblco phase diagram} and~\ref{compare}). We can clearly see that
the rate at which $\nu_{Q}a^{3}$ increases with increasing $y$ varies between
the two families. The data from the different families generate two different
linear curves. Two straight lines are fitted to these datasets with the
constraint that the slopes ratio is $1.62$, which, as mentioned before, is the
ratio of $K(0.4)$ to $K(0.1)$. The measured $\nu_{Q}a^{3}$ versus $y$ can be
explained well by the two lines. When we perform a linear fits with no
constrains the slopes ratio is 2.2$\pm$0.65 which is within error bars equal
to the $K$'s ratio. This is the main experimental finding of this work.%

Since we have demonstrated experimentally that $\Delta(\nu_{Q}a^{3})\propto
K(x)\left(  y-y_{N}\right)  $ and argued above that $\Delta(\nu_{Q}%
a^{3})\propto\Delta n_{p_{\sigma}}$, we conclude that%
\begin{equation}
\Delta n_{p_{\sigma}}\propto K(x)\left(  y-y_{N}\right)  . \label{Stretching1}%
\end{equation}
Therefore, the doping efficiency of the $p_{\sigma}$ orbital is family
dependent. We would like to emphasize again that we can only quantify the
doping efficiency ratio between the two different families ($K$s ratio), but
not their absolute value. Hence the proportionality sign in
Eq.~\ref{Stretching1}. The set of $K$s which generate Fig.
\ref{clblco phase diagram}(b) are chosen to give $\Delta n_{p_{\sigma}}=0.05$
at optimal doping according to Hasse \textit{et al.} \cite{Hasse}\textbf{.
}Negative values of $\Delta n_{p_{\sigma}}$ represent CuO$_{2}$ planes which
are not doped. Another degree of freedom is the number of $p_{\sigma}$ holes
for $y<y_{N}$. Following Ref. \cite{Hasse} again, by setting $n_{p_{\sigma}%
}(y<y_{N})=0.11$ we obtain $n_{p_{\sigma}}=0.16$ at optimal doping.

There were several attempts in the past to find a relation between the number
of holes in the CuO$_{2}$ plane and oxygen level $y$ of CLBLCO. Chmaissem
\emph{et al.} used bond valence summation (BVS) calculations based on
structural parameters determined by neutron diffraction \cite{Chmaissem}.
Keren \emph{et al.} measured the in plane $^{63}$Cu NQR parameter $^{63}%
\nu_{Q}$ which is shown in the inset of Fig.~\ref{slopes}, and Sanna \emph{et
al.} experimented with x-ray fine structure (XFS) \cite{Sanna}. BVS has some
theoretical arbitrariness and is not completely reliable. $^{63}\nu_{Q}(y)$
shows no family dependence because $^{63}\nu_{Q}$ is sensitive to charge on
the apical O$_{4}$ $p_{\sigma}$, Cu $3d_{x^{2}-y^{2}},$ $3d_{z^{2}-r^{2}}$ and
$4s$, and O$_{2,3}$ $p_{\sigma}$ holes simultaneously \cite{Hasse}; hence it
is not an ideal probe and a difference in the slopes of $^{63}\nu_{Q}(x,y)$
could not be detected within the experimental error bars. Finally, the XFS
peak is constructed of three contributions \cite{Sanna} which again limit
their resolution. Therefore, none of the three attempts could find a
difference in the doping efficiency of the planes within experimental
resolution. The oxygen NQR has the advantage of measuring directly the
dependence of $p_{\sigma}$ hole density $n_{p\sigma}$ on the oxygen level $y$,
and, indeed, this probe detects variations in doping efficiency.

The physical meaning of Fig. \ref{slopes} and Eq.~\ref{Stretching1} is that
the efficiency of the doping process, namely the injection of holes into the
oxygen $p_{\sigma}$ orbital, varies between the CLBLCO families. Moreover, the
scaling procedure leading from Fig.~\ref{clblco phase diagram}(a) to Fig.~
\ref{clblco phase diagram}(b) can now be fully justified: the third step is
needed because the oxygen density is not the relevant parameter and one must
use $n_{p_{\sigma}}$ in the phase diagram. The second step means that
$T_{c}^{\max}$ is determined by the magnetic super-exchange interaction energy
scale $J$, namely, $J(x)/T_{c}^{\max}(x)$ is $x$ independent. This leads to
the unified phase diagram of Fig.~\ref{clblco phase diagram}(b) which is obtained
with no adjustable parameters.

This work was funded by the Israeli Science Foundation.

\end{document}